\documentclass[aps,prb,superscriptaddress,twocolumn,showpacs,amsmath,amssymb]{revtex4}
\usepackage{graphicx}
\usepackage{epsfig}
\usepackage{pslatex}
\usepackage{natbib}

\newcommand{\chem}[1]{\ensuremath{\mathrm{#1}}}

\begin{document}

\title{Probing magnetic order in Li{\it M}PO$_{4}$, $M=$~Ni, Co, Fe and lithium diffusion in Li$_x$FePO$_{4}$}

\author{P. J. Baker}
\email{peter.baker@stfc.ac.uk}
\affiliation{ISIS Facility, STFC Rutherford Appleton Laboratory, Chilton, Oxon. OX11 0QX. United Kingdom}
\author{I.~Franke}
\affiliation{Clarendon Laboratory, University of Oxford, Parks Road, Oxford. OX1 3PU. United Kingdom}
\author{F.~L.~Pratt}
\affiliation{ISIS Facility, STFC Rutherford Appleton Laboratory, Chilton, Oxon. OX11 0QX. United Kingdom}
\author{T.~Lancaster}
\affiliation{Clarendon Laboratory, University of Oxford, Parks Road, Oxford. OX1 3PU. United Kingdom}
\author{D.~Prabhakaran}
\affiliation{Clarendon Laboratory, University of Oxford, Parks Road, Oxford. OX1 3PU. United Kingdom}
\author{W.~Hayes}
\affiliation{Clarendon Laboratory, University of Oxford, Parks Road, Oxford. OX1 3PU. United Kingdom}
\author{S.~J.~Blundell}
\affiliation{Clarendon Laboratory, University of Oxford, Parks Road, Oxford. OX1 3PU. United Kingdom}
\date{\today}

\begin{abstract}
Muon spin relaxation measurements are reported on three members of the Li$_x${\it M}PO$_4$ series. The magnetic properties of 
stoichiometric samples with $M=$~Ni, Co, Fe, were investigated at low-temperature.
In LiNiPO$_4$ we observe different forms of the muon decay asymmetry in the commensurate and incommensurate antiferromagnetic 
phases, accompanied by a change in the temperature dependence of the muon oscillation frequency.
In LiCoPO$_4$ the form of the muon decay asymmetry indicates that the correlation between layers decreases as the N\'{e}el 
temperature is approached from below. 
LiFePO$_4$ shows more conventional behaviour, typical for an antiferromagnet. 
Measurements on Li$_x$FePO$_4$ with $x = 0.8,~0.9$~\&~$1$ show evidence for lithium diffusion below $\sim 250$~K and muon 
diffusion dominating the form of the relaxation at higher temperature. The thermally activated form of the observed hopping rate 
suggests an activation barrier for lithium diffusion of $\sim 100$~meV and a diffusion constant of $D_{\rm Li} \sim 10^{-10} - 
10^{-9}$~cm$^2$s$^{-1}$ at room temperature.
\end{abstract}


\maketitle

\section{\label{sec:intro} Introduction}
The series of phosphates Li{\it M}PO$_4$ crystallize in the orthorhombic olivine structure, with layers of magnetic transition 
metal {\it M} ions that are relatively well coupled, meaning they are intermediate between two- and three-dimensional 
magnetism.~\cite{geller60,vakninprl} The choice of {\it M} ion allows the single ion interactions to be tuned discretely and a 
range of magnetic behaviour results.
In LiNiPO$_4$ the low-temperature commensurate antiferromagnetic state becomes incommensurate just below the bulk ordering 
temperature.~\cite{vakninprl}
Both LiCoPO$_4$ and LiNiPO$_4$ exhibit magnetoelectric behaviour~\cite{kornev00} and the resulting toroidal domains in LiCoPO$_4$ 
have been observed using optical measurements.~\cite{vanaken}

A separate interest in this series of phosphates comes from their application as battery cathode materials. This is particularly 
relevant for LiFePO$_4$ which has a slightly lower cell voltage and energy density than the widely used LiCoO$_2$, but a 
significantly better lifetime, resistance to thermal runaway, and a smaller environmental 
impact.~\cite{padhi97,yamada,whittingham05,ohzuku07,yuan11}
The use of Li$_x$FePO$_4$ as a battery cathode material leads to questions concerning its electrochemical properties and the 
kinetics of lithium diffusion, both of which have received considerable study.~\cite{yuan11} Both calculations and experiment have 
addressed the activation barriers for lithium ion and electron conduction, as well as the lithium ion diffusion 
rate.~\cite{ellis06,dodd07,prosini02,churikov10,franger02,yu07,morgan04,islam,amin08,takahashi02,wang07,maxisch06,hoangxx,cabana10
} While the calculated values have converged there are considerable variations in the results of experiments carried out using 
different techniques.

In this paper we present a muon-spin relaxation ($\mu$SR) investigation of the low-temperature magnetic properties of Li{\it 
M}PO$_{4}$ ($M=$~Ni, Co, Fe) and the high-temperature diffusive properties of Li$_x$FePO$_{4}$ ($x=$~0.8, 0.9, 1.0). As well as 
being a sensitive probe of magnetic ordering,~\cite{blundell99,yddrbook} $\mu$SR provides a means of investigating diffusion 
processes of both the muon~\cite{kadono89} and other species that perturb its environment.~\cite{sugiyama09} Lithium diffusion is 
a process that provides such a perturbation and $\mu$SR has now been applied to studying it in a wide range of battery cathode 
materials: 
Li$_x$[Mn$_{1.96}$Li$_{0.04}$]O$_4$,~\cite{kaiser00,ariza03} Li$_{0.6}$TiO$_2$,~\cite{gubbens06} 
Li$_{3-x-y}$Ni$_x$N,~\cite{powell09}, Li$_x$CoO$_2$,~\cite{mukai07,sugiyama09} and LiNiO$_2$.~\cite{sugiyama10} Of these, the 
studies on Li$_x$CoO$_2$ have been the most extensive and found that the lithium diffusion rate in this compound is well suited to 
the timescale probed by $\mu$SR. Similar studies on a different timescale can be carried out using 
NMR.~\cite{verhoeven01,nakamura06,cabana10}

We describe the preparation of the samples and the general $\mu$SR technique in section~\ref{sec:exp}, including the details of 
how the low-temperature data were analysed. In sections~\ref{sec:lini}, \ref{sec:lico}, and~\ref{sec:life} we present our $\mu$SR 
data and analysis. In section~\ref{sec:diff} we discuss the existing literature on lithium diffusion in Li$_x$FePO$_4$, describe 
our higher-temperature $\mu$SR experiments, data analysis, and results. Our conclusions are summarized in section~\ref{sec:conc}.

\section{\label{sec:exp} Experimental Method}
Powders of stoichiometric and Li-deficient Li$_x${\it M}PO$_4$ ($M=$Ni, Co and Fe; $x=0.8$ and $0.9$) were synthesized by the 
solid state reaction technique.  Starting materials of high purity ($>99.99$~\%) Li$_3$PO$_4$, NiO, Co$_3$O$_4$, Fe$_2$O$_3$, and 
NH$_4$H$_2$PO$_4$ were mixed and sintered in three stages; $175^{\circ}$~C for 10~h, $225^{\circ}$~C for 5~h, and $725^{\circ}$~C 
for 24~h.   After grinding, they were sintered again at $750^{\circ}$~C for 24~h. Finally the powders were made into rods and 
sintered at $775^{\circ}$~C for 12~h. Single crystals of stoichiometric Li{\it M}PO$_4$ were grown in a four-mirror optical 
floating-zone furnace (Crystal System Inc.).  The growth was carried out at a speed of 2--3~mm/h with the feed and seed rods 
counter rotating at 25~rpm.  Crystals were grown in an argon pressure of 1--4 atmosphere.  

Spin-polarized positive muons were implanted into the samples where they stop at interstitial sites with large electronegativity, 
and decay with a mean lifetime of $2.2~\mu$s. While the muons remain within the sample their spin direction is affected by the 
local magnetic field at their stopping site, with the muon's gyromagnetic ratio $\gamma_{\mu}=2\pi \times 135.5$~MHz~T$^{-1}$ 
being intermediate between those of the electron and proton. 
The muon spin polarization is followed as a function of time by measuring the asymmetry in the count rate of decay positrons, 
$A(t)$, in two detectors on opposite sides of the sample.~\cite{blundell99}
Our $\mu$SR experiments were carried out at the Paul Scherrer Institute using the General Purpose Surface-muon instrument (GPS) 
for the low-temperature measurements of the stoichiometric crystalline samples and at the ISIS Facility using the MuSR 
spectrometer for the high-temperature measurements of the Li$_x$FePO$_{4}$ samples.

\begin{figure}[t]
\includegraphics[width=\columnwidth]{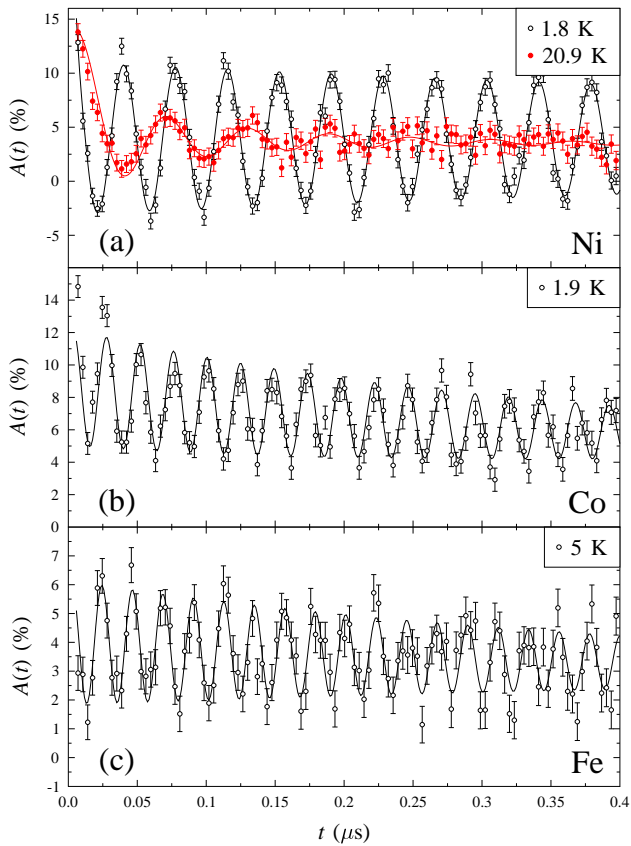}
\caption{\label{raw} (Color online) 
Muon decay asymmetry data for \chem{Li{\it M}PO_{4}}: (a) $M=$Ni, (b) $M=$Co, and (c) $M=$Fe. 
Fits to the data are described in the text. The initial asymmetries are reduced from the values due to energy selection of decay 
positrons by including data from spatially segmented detectors giving better counting statistics.
}
\end{figure}

The data shown in Fig.~\ref{raw} were analysed using the WiMDA program.~\cite{wimda} It was found that similar fitting functions 
were suitable for describing the data on each of the three stoichiometric samples, based on the general form:
\begin{equation}
A(t) = A_{\perp} e^{-\lambda t}\cos(2\pi f t + \phi) + A_{\parallel} e^{-\Lambda t} + A_{\rm bg}.
\label{fitfunc}
\end{equation}
The $A_{\perp}$ term describes an exponentially damped oscillation due to a quasistatic magnetic field perpendicular to the muon 
spin direction. The phase $\phi$ could be fixed to zero in both the low-temperature commensurate phase of LiNiPO$_4$, and below 
$T_{\rm N}$ in LiFePO$_4$. In LiCoPO$_4$, $\phi$ was found to depend on temperature. The $A_{\parallel}$ term describes the 
exponential relaxation for muon spins with their direction along that of the local field at their stopping site, which are 
depolarized by spin fluctuations. The final term describes the temperature-independent contribution to the asymmetry from muons 
stopping outside the sample. Just above the magnetic ordering transition there is no oscillatory signal and, as is generally the 
case in paramagnets, the data are well described by an exponential relaxation, with rate $\Lambda$. This form of the data is not 
compatible with short-ranged static magnetic order persisting above the long-range magnetic ordering transition on the timescales 
probed by muons. We discuss the form of the muon depolarization at higher temperatures in section~\ref{sec:diff}.

In the incommensurate antiferromagnetic phase of LiNiPO$_4$ we found that Eq.~\ref{fitfunc} did not provide a satisfactory 
description of the data, even allowing $\phi$ to change from the value of zero found to describe the commensurate phase. For 
incommensurate magnetic phases, where the muons sample a significant range of magnetic fields, the oscillatory part of the muon 
relaxation function takes the form of a Bessel function,~\cite{yddrbook} so that $A(t)$ can be written as:
\begin{equation}
A(t) = A_{\perp} e^{-\lambda^{\prime}t}J_0(2\pi f^{\prime}t)+ A_{\parallel} e^{-\Lambda t} + A_{\rm bg}.
\label{besselfunc}
\end{equation}

For well-defined oscillation frequencies that varied continuously below $T_{\rm N}$ we fitted the temperature dependence to the 
phenomenological function: 
\begin{equation}
f(T) = f(0) [1-(T/T_{\rm N})^{\alpha}]^{\beta},
\label{nuT}
\end{equation}
where $\alpha$ describes the $T \rightarrow 0$ trend and $\beta$ describes the trend approaching $T_{\rm N}$. 

\section{\label{sec:results} Low-temperature results}
\subsection{\label{sec:lini} \chem{LiNiPO_4}}

\begin{figure}
\includegraphics[width=\columnwidth]{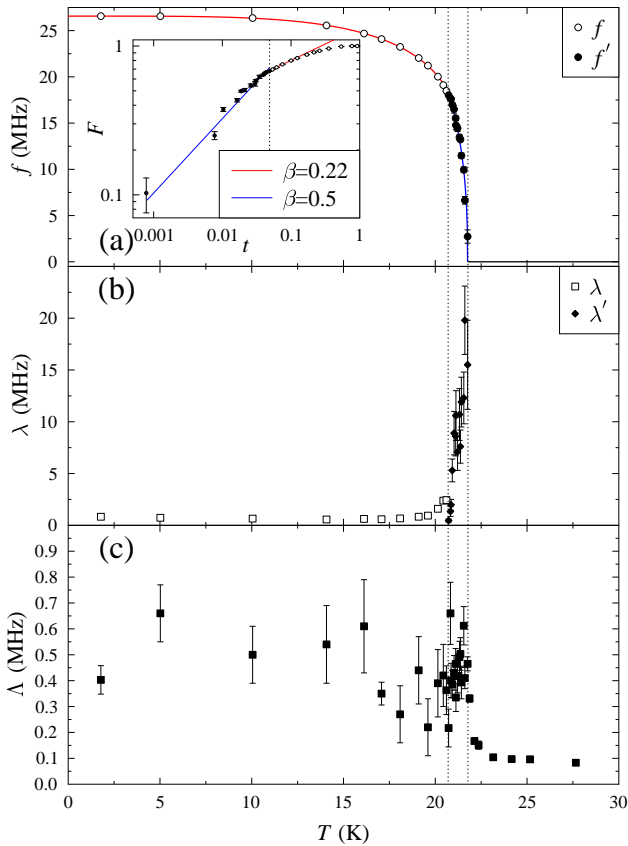}
\caption{\label{lini} (Color online)
Parameters derived from fitting equations~\ref{fitfunc} and~\ref{besselfunc} to the $\mu$SR data for LiNiPO$_4$ shown in 
Fig.~\ref{raw}~(a): 
(a) Oscillation frequencies $f$ and $f^{\prime}$.
(Inset) Reduced oscillation frequency vs reduced temperature (see text) showing the kink at the commensurate-incommensurate phase 
transition, with guides to the eye plotted for each of the two phases.
(b) Linewidths $\lambda$ and $\lambda^{\prime}$.
(c) Relaxation rate $\Lambda$.
}
\end{figure}

On cooling LiNiPO$_4$ orders incommensurately at $T_{\rm N} = 21.8$~K, then orders commensurately at $T_{\rm IC-C} = 
20.7$~K.~\cite{vakninprl,kharchenko03} Neutron diffraction has characterized both the long-range ordered phases and found diffuse 
scattering well above $T_{\rm N}$. The trend in the magnetic order parameter approaching the breakup of collinear order was found 
to follow the dependence expected for a 2D Ising model;~\cite{vakninprbni} an anomalous correlation between the spin wave spectrum 
and the incommensurate magnetic order~\cite{jensenprb413} and coexisting short- and long-range incommensurate magnetic order were 
reported for the intermediate phase.~\cite{vakninprl} 

In LiNiPO$_4$ we can divide the data into three distinct temperature regions. At the lowest temperatures, below $20.7$~K, we 
observe underdamped oscillations at a single frequency, with the data shown in Fig.~\ref{raw}~(a) well described by 
Equation~\ref{fitfunc} with $\phi=0$. This is consistent with the commensurate magnetic structure determined by neutron 
diffraction.~\cite{vakninprl} The temperature dependence of the oscillation frequency, $f$, is shown in Fig.~\ref{lini}~(a), the 
linewidth $\lambda$ in (b), and the relaxation rate $\Lambda$ in (c). The linewidth is relatively small compared to the 
oscillation frequency, as is evident from the persistence of the oscillations in the low-temperature data, and only grows slightly 
approaching the commensurate-incommensurate phase transition.

Between $20.7$ and $21.8$~K the oscillations persist, but their form changes from the cosinusoidal form described by 
equation~\ref{fitfunc} with $\phi = 0$ to the Bessel function form described by equation~\ref{besselfunc} as expected from the 
incommensurate behaviour determined by neutron diffraction.~\cite{vakninprl} While it is possible to obtain convergent fits to 
Eq.~\ref{fitfunc} with $\phi$ as a free parameter the quality of fit is markedly poorer than for Eq.~\ref{besselfunc} and the 
difference is obvious even to the eye. The oscillation frequency, $f^{\prime}$, linewidth, $\lambda^{\prime}$, and relaxation rate 
$\Lambda$ are shown in Fig.~\ref{lini}~(a), (b), and (c) respectively. The linewidth grows more rapidly in this temperature region 
which suggests that it is dominated by critical fluctuations approaching $T_{\rm N}$.

Fitting the oscillation frequencies shown in Fig.~\ref{lini}~(a) using Eq.~\ref{nuT}, extended to two phases with different 
$\beta$ values in each phase but a continuous order parameter, leads to the parameters: $T_{\rm N} = 21.76(1)$~K, 
$f(0)=26.57(1)$~MHz, $\alpha=4.22(5)$, $\beta_{\rm C} = 0.220(3)$, and $\beta_{\rm IC} = 0.40(5)$, the last two parameters being 
in the commensurate (C) and incommensurate (IC) ordered phases respectively.
The relaxation rate $\Lambda$ does not display any clear trend in the lowest temperature phase but there is a distinct rise in the 
incommensurate phase. Above $T_{\rm N}$ the data take the exponential form typical of paramagnets where the electronic spin 
fluctuations are fast compared to the distribution of local fields. The relaxation rate also appears to vary critically above 
$T_{\rm N}$.

In the inset to Fig.~\ref{lini}~(a) we show the reduced oscillation frequencies plotted against reduced temperature, taking 
$f(T=0)$ and $T_{\rm N}$ as the fixed points for $F = f(T)/f(0)$ and $t=(T_{\rm N}-T)/T_{\rm N}$ respectively. Trends following 
the expected $\beta$ values for the two phases are sketched as guides to the eye, without the $\alpha$ parameter used in fitting 
the trend shown in the main panel. A kink in the plot of $F$ against $t$ is evident at the commensurate-incommensurate phase 
transition.

\subsection{\label{sec:lico} \chem{LiCoPO_4}}

\begin{figure}[t]
\includegraphics[width=\columnwidth]{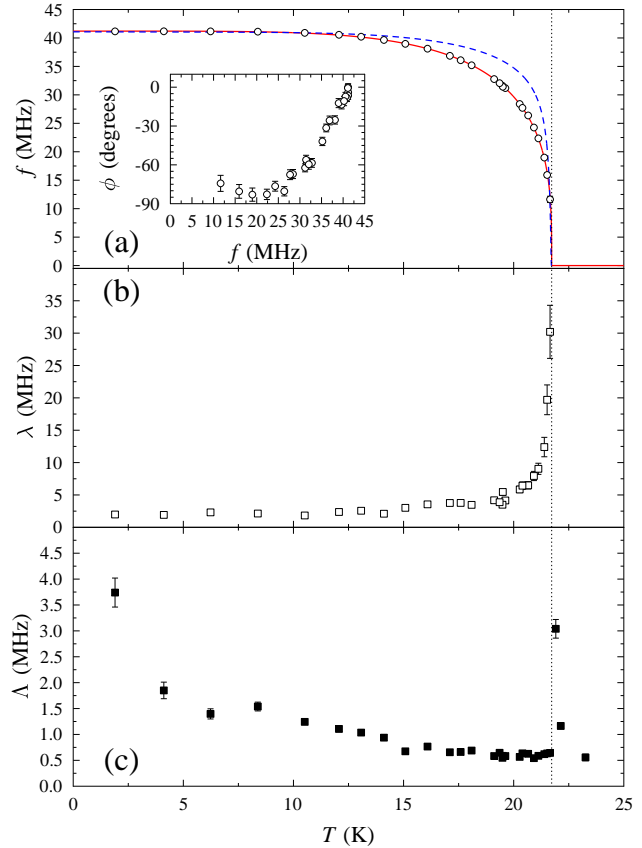}
\caption{\label{lico} (Color online)
Parameters derived from fitting Eq.~\ref{fitfunc} to the $\mu$SR data for LiCoPO$_4$ shown in Fig.~\ref{raw}~(c). 
(a) Oscillation frequency $f$ with the solid line showing a fit to Eq.~\ref{nuT} and the dashed line showing the model derived in 
Ref.~\onlinecite{vakninprbco}.
(Inset) Phase of oscillating component $\phi$ plotted against the oscillation frequency $f$. This shows that they do not have a 
monotonic relation, excluding the possibility that a time offset is causing the phase offset.
(b) Linewidth $\lambda$.
(c) Relaxation rate $\Lambda$.
}
\end{figure}

LiCoPO$_4$ has been found to be a model example of a magnet intermediate between 2D and 3D exchange coupling.~\cite{vakninprbco} 
This gives rise to some unusual critical behaviour, with a neutron diffraction study reporting that the temperature dependence of 
the staggered magnetization follows the form expected for the 2D Ising model and that the critical scattering above $T_{\rm N} = 
22$~K follows a 3D Ising form.~\cite{vakninprbco} The spin waves have also been studied in detail~\cite{tian}, showing that as 
well as the dispersions predicted by linear spin wave theory, there is an anomalous dispersionless excitation at $\sim 1.2$~meV 
that was suggested to be related to the magnetoelectric effect in this material.

The form of the raw data for LiCoPO$_4$, shown in Fig.~\ref{raw}~(b), is similar to that in the commensurate phase of LiNiPO$_4$, 
and can be fitted successfully using Eq.~\ref{fitfunc}. However, it was not possible to constrain $\phi=0$ all the way up to 
$T_{\rm N}$. In such a situation, caution is required in distinguishing between a systematic shift in the time offset of the raw 
data (arising from an error in determining when muons enter the sample) and a phase offset due to the magnetic field distribution 
in the sample. Both possibilities were considered in the data analysis. Because the muons were implanted into the sample with 
their initial spin direction rotated relative to the symmetry axes of the detector system it was possible to use the geometric 
phase offsets to show that only the phase of the signal was varying with temperature. The data were subsequently analyzed with 
$\phi$ as a free parameter, leading to the parameters shown in Fig.~\ref{lico}.

The oscillation frequency shows a smooth, monotonic decrease with increasing temperature and can be fitted to the empirical form 
of Eq.~\ref{nuT} with the parameters: $f(0)=41.20(1)$~MHz, $T_{\rm N} = 21.72(1)$~K, $\alpha = 4.91(4)$, and $\beta = 0.299(3)$. 
This value of $\beta$ is smaller than that expected for the 3D Ising model ($\beta = 0.326$), but considerably larger than that 
for the 2D Ising model ($\beta=0.125$).
Vaknin {\em et al.}~\cite{vakninprbco} compared the results of their neutron diffraction measurements to the analytic Onsager/Yang 
form~\cite{onsageryang} $M^{\dag}(T) = M^{\dag}(0)[1-\sinh^{-4}(2J_{2D}/T)]^{1/8}$ for the temperature dependence of the 
sublattice magnetization in the 2D Ising model, including an additional multiplicative term $\exp[\Delta_{\rm E - G}/(T-T_{\rm 
N})]$ to describe the interlayer fluctuations relevant near the crossover to 3D behaviour. 
Neither the purely 2D model nor the extended version were able to describe $f(T)$ above $15$~K, although the additional 
multiplicative term does bring the predicted order parameter closer to the trend we observe. We plot the coupled layer model of 
Ref.~\onlinecite{vakninprbco} as a dashed line in Fig.~\ref{lico}~(a) to illustrate this difference. Given the effectiveness of 
the analytic model below $15$~K and the unusually large value of $\alpha$ (dominated by these low-temperature data points), it 
seems that there is a gradual crossover in the effective dimensionality of the system around $15$~K that is reflected in the form 
of the muon data. 
This is consistent with the 3D fluctuations occurring on the longer timescale probed by muons at a lower temperature than they 
affect neutron diffraction measurements.

The linewidth $\lambda$ shown in Fig.~\ref{lico}~(b) is slightly larger than in LiNiPO$_4$ at low temperature and appears to grow 
in two stages as $T_{\rm N}$ is approached: around $15$~K there is a small step and above $20$~K there is a sharper rise 
associated with critical fluctuations. The relaxation rate $\Lambda$ shows almost the opposite temperature dependence to $\lambda$ 
and is considerably larger than in either of the other two samples. There is no sign of a critical divergence approaching $T_{\rm 
N}$ from below and these observations suggest that $\Lambda$ is dominated by a quasistatic distribution of magnetic fields. Above 
$T_{\rm N}$, $\Lambda$ shows a more conventional critical divergence.

The temperature dependence of the phase $\phi$ is plotted in the inset of Fig.~\ref{lico}~(a). Below $15$~K there is only a slow 
change in $\phi$ but it rapidly changes from $-10^{\circ}$ to $\sim -80^{\circ}$ above $15$~K.
This suggests that, approaching $T_{\rm N}$, a magnetic inequivalency develops between muon stopping sites in a manner akin to the 
intermediate phase in LiNiPO$_4$, albeit less pronounced. 
While a weakly incommensurate structure could generate such an effect, the sharp increase in $\phi$ occurring at the same 
temperature as both the departure from the quasi-2D trend in $f(T)$ and the growth in the linewidth suggests that in LiCoPO$_4$ 
the phase shift comes from increasing disorder in the stacking of the magnetic layers while approaching $T_{\rm N}$ from below.

\subsection{\label{sec:life} \chem{LiFePO_4}}
\begin{figure}[t]
\includegraphics[width=\columnwidth]{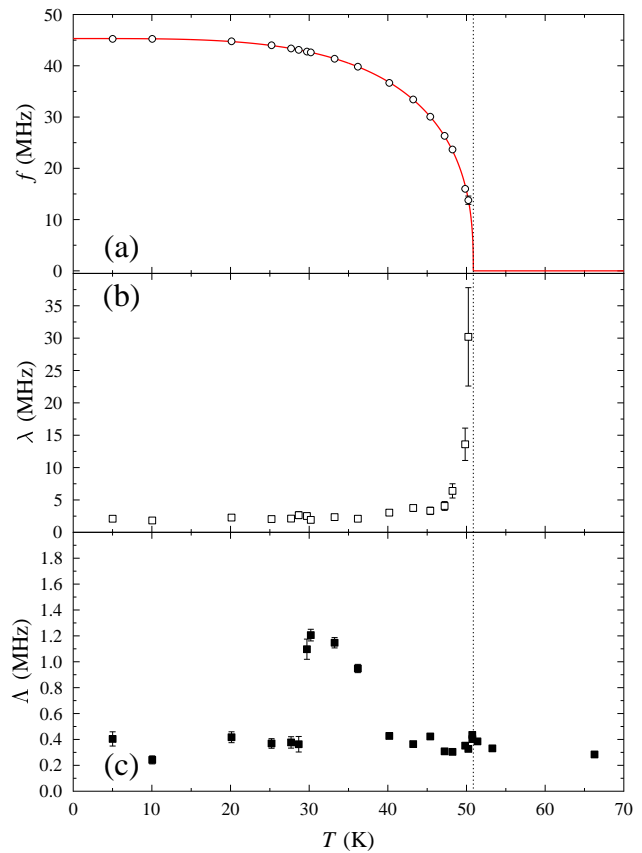}
\caption{\label{life} (Color online)
Parameters derived from fitting Eq.~\ref{fitfunc} to the $\mu$SR data for LiFePO$_4$ shown in Fig.~\ref{raw}~(b). 
(a) Oscillation frequency $f$.
(b) Linewidth $\lambda$.
(c) Relaxation rate $\Lambda$.
}
\end{figure}

LiFePO$_4$ shows a commensurate collinear antiferromagnetic structure below $T_{\rm N} = 50$~K, with neutron diffraction 
measurements finding $\beta = 0.27(3)$.~\cite{rousse,li06} The Fe moments are orientated slightly away from the $b$-axis, expected 
on the basis of the crystal symmetry. No evidence for short-range order has been observed above $T_{\rm N}$, in contrast to the 
other members of this series. 

The parameters derived from fits of equation~\ref{fitfunc} to the LiFePO$_4$ data are shown in Fig.~\ref{life}. The temperature 
dependence of the frequency and linewidth behave conventionally, with $f(T)$ being well described by equation~\ref{nuT} with 
parameters: $f(0)=45.31(2)$~MHz, $T_{\rm N} = 50.87(7)$~K, $\alpha = 3.66(3)$, and $\beta = 0.381(5)$. This value of $\beta$ is 
close to that expected for 3D Heisenberg critical behaviour, as opposed to the value of $\beta = 0.27(3)$ previously estimated 
from neutron scattering measurements.~\cite{rousse,li06} The linewidth for $T \ll T_{\rm N}$ is considerably smaller than the 
oscillation frequency and it diverges smoothly approaching $T_{\rm N}$.

The relaxation rate $\Lambda$ behaves differently in LiFePO$_4$, compared with the other two samples. There is only a small 
increase around $T_{\rm N}$ but the primary feature occurs at $30$~K, well below $T_{\rm N}$. 
Associated with this feature is an increase in the relaxing amplitude of the signal as the temperature is increased. Examining the 
low-temperature data more carefully allows us to identify a further oscillating component with an amplitude around $10$~\% of the 
primary oscillating component, with an oscillation frequency of $\sim 120$~MHz. This component disappears above $30$~K and appears 
to be replaced by the strongly relaxing term that causes an increase in $\Lambda$. We attribute this behavior to a $\sim 10$~\% 
impurity phase that is most likely to be FePO$_4$, since Li$_x$FePO$_4$ is known to form an $x$LiFePO$_4$:$(1-x)$FePO$_4$ binary 
phase mixture~\cite{yuan11} and $T_{\rm N}$(FePO$_4$)$\simeq 25$~K.~\cite{battle82}

\section{\label{sec:diff} High-temperature results for \chem{Li_{\it x}FePO_4}}
\begin{figure}[t]
\includegraphics[width=\columnwidth]{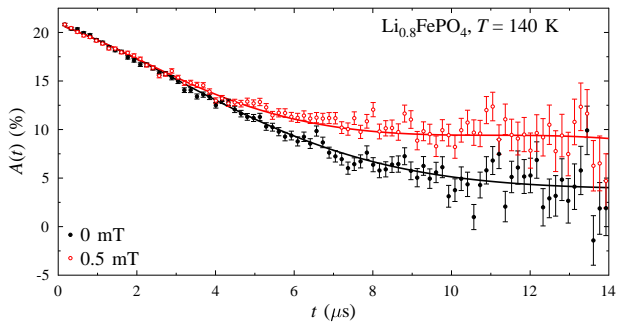}
\caption{\label{hiT} (Color online)
Raw $\mu$SR data for Li$_{0.8}$FePO$_4$ at 140~K with fits to Eq.~\ref{keren1} as described in the text.
}
\end{figure}

\begin{figure}[t]
\includegraphics[width=\columnwidth]{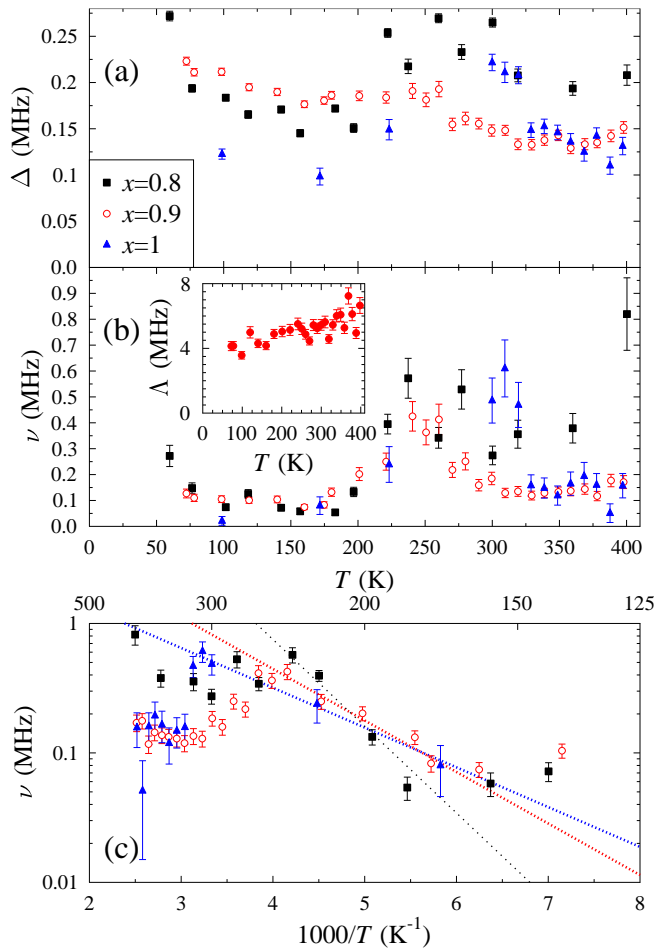}
\caption{\label{diff} (Color online)
Parameters derived from fitting Eq.~\ref{keren1} to the $\mu$SR data for Li$_x$FePO$_4$. 
(a) Field distribution width $\Delta$.
(b) Fluctuation rate $\nu$.
(Inset) Relaxation rate $\Lambda$ for Li$_{0.9}$FePO$_4$.
(c) Fluctuation rate $\nu$ plotted against the inverse temperature to illustrate the thermally activated behavior below about 
250~K. The lines plotted are fits to the data in the activated region with parameters described in the text.
}
\end{figure}

To investigate the lithium diffusion behavior in Li$_x$FePO$_4$ we measured three samples with $x=0.8$, $0.9$, and $1.0$ at 
temperatures between $75$ and $400$~K and at fields of $0$ and $0.5$~mT. By measuring at more than one magnetic field at each 
temperature it is possible to get a more reliable determination of the fitted parameters, since we have more information on how 
the field distribution experienced by the muon is decoupled by the field applied parallel to the initial muon spin polarization. 
Examples of the raw data at the two magnetic fields used are shown in Fig.~\ref{hiT} with the fits described below.

For our high-temperature measurements on Li$_x$FePO$_4$ we assume a Gaussian distribution of random local fields due to the 
various magnetic moments present in the sample. For a static magnetic system this would lead to a muon depolarization described by 
the Gaussian Kubo-Toyabe function.~\cite{hayano79} Any fluctuations present within the muon time window, which may be caused by 
either lithium or muon diffusion, can be treated using the strong collision approximation, leading to a dynamic Kubo-Toyabe 
function.~\cite{hayano79} Analysis of the data measured at a series of fields and temperatures using such a dynamic Kubo-Toyabe 
function proved to be unsuccessful. In studies of lithium-containing battery materials it has been usual to multiply the dynamic 
Gaussian Kubo-Toyabe function by an exponential relaxation to eliminate any magnetic contribution to the 
relaxation.~\cite{sugiyama09,kaiser00,powell09,sugiyama10} 
This does not lead to reliable fits to our raw data either. A consistently better quality of fit was obtained by applying Keren's 
analytic generalization of the Abragam function appropriate for $\mu$SR,~\cite{keren} multiplied by a temperature-independent 
relaxation rate fixed for each sample:
\begin{equation}
P_{z}(t) = \exp[-\Gamma(\Delta,\nu,\omega_{L},t)t]\exp(-\lambda t).
\label{keren1}
\end{equation}
where $\Gamma(\Delta,\nu,\omega_{L},t)$ is defined in Ref.~\onlinecite{keren}. The parameter $\Delta$ describes the quasistatic 
distribution of field at the muon stopping site, $\nu$ is the temperature-dependent fluctuation rate, $\omega_L = \gamma_{\mu}B$ 
is the muon's Larmor precession frequency in the applied magnetic field, and $\lambda$ is due to temperature-independent 
fluctuations. (After initial unconstrained fits had been made, $\lambda$ values were fixed at $0.05$, $0.02$, and $0.1$~MHz for 
the $x=0.8$, $0.9$, and $1.0$ samples respectively.) In the $x=0.9$ sample we found a strong temperature independent relaxation 
coming from a minority phase which could be subtracted from the data analysis using $A_{\rm m} \exp(-\Lambda t)$, with the values 
of $\Lambda$ shown in the inset to Fig.~\ref{diff}~(b). The values of $\Delta$ and $\nu$ obtained from these fits are shown in 
Fig.~\ref{diff}~(a) and (b) respectively. 

The $\Delta$ values in Fig.~\ref{diff}~(a) show the trend observed in the vast majority of lithium-containing battery materials 
investigated to date for $x=0.9$, where a low-temperature plateau is followed by a smooth decrease to a high-temperature 
plateau.~\cite{sugiyama09,kaiser00,powell09,sugiyama10} In the $x=0.8$ and $1$ samples there is a peak at around the temperature 
where the low-temperature plateau ends in the $x=0.9$ sample.
The values of $\Delta \sim 0.2$~MHz are broadly similar to those in Li$_x$CoO$_2$~\cite{sugiyama09}, 
LiMn$_2$O$_4$,~\cite{kaiser00} and Li$_{1-x}$Ni$_{1+x}$O$_2$,~\cite{sugiyama10} but smaller than in 
Li$_{3-x-y}$Ni$_x$N.~\cite{powell09}

The temperature dependence of $\nu$ follows a similar trend in each sample, with a slight fall from the lowest measured 
temperature to around $100$~K, followed by a smooth rise towards $\sim 250$~K, and then a sharp drop-off to either the 
low-temperature value, or in $x=0.8$, to the value at the peak. It seems likely that the change observed below 100~K is due to the 
buildup of magnetic correlations that are not well described by our temperature-independent $\lambda$ value. Above 100~K the 
thermally activated growth in the hopping rate mirrors that observed in Li$_x$CoO$_2$~\cite{sugiyama09} and 
Li$_{1-x}$Ni$_{1+x}$O$_2$,~\cite{sugiyama10} albeit with a different temperature scale. The behavior above the peak at $\sim 
250$~K may be related to the onset of muon hopping, but this may not be a unique explanation.

\begin{table}[t]
\caption{\label{table}
Comparison of reported estimates for $D_{\rm Li}$ and $E_a$ obtained using different techniques (at room temperature unless 
noted). A more detailed list of $E_a$ values is given in Ref.~\onlinecite{amin08}.
}
\begin{ruledtabular}
\begin{tabular}{lcc}
Technique & $D_{\rm Li}$ (cm$^2$s$^{-1}$) & $E_a$ (meV) \\
\hline
$\mu$SR (This study) & $10^{-10} - 10^{-9}$ & $\sim 100$ \\
M\"{o}ssbauer spectroscopy~\cite{ellis06} & $10^{-7}$ & $775 \pm 108$\footnote{Determined around 600~K.} \\
M\"{o}ssbauer spectroscopy~\cite{dodd07} & $10^{-13} - 10^{-11}$ & $335 \pm 25$\footnote{Determined around 450~K.} \\
Titration and ac impedance~\cite{prosini02} & $10^{-15}$ & - \\
Titration~\cite{churikov10} & $10^{-16} - 10^{-10}$ & - \\
Impedance~\cite{franger02} & $10^{-14}$ & - \\
Cyclic voltammetry~\cite{yu07} & $10^{-14}$ & 400 \\
First-principles calculations~\cite{morgan04} & $10^{-8}$ & 270 \\
First-principles calculation~\cite{islam} & - & 550 \\
AC and DC conductivity~\cite{amin08} & - & 620 -- 740 \\
AC impedance~\cite{takahashi02} & - & 155 \\
Electrochemistry\cite{wang07} & - & 155 \\
\end{tabular}
\end{ruledtabular}
\end{table}

Comparing our results to those obtained previously naturally leads to the question of whether the phenomena we observe are 
associated with the diffusion of lithium and/or muons. The similarity of the temperature dependences of both $\Delta$ and $\nu$ to 
previous results on other materials indeed suggest that they are caused by the same phenomenon. This leaves the further question 
of whether we can obtain quantitative information about the lithium diffusion from our results. While we could not use the dynamic 
Gaussian Kubo-Toyabe function multiplied by an exponential used in Refs.~\onlinecite{sugiyama09},~\onlinecite{kaiser00}, 
and~\onlinecite{sugiyama10}, the modified Keren function we have employed provides the same information and a more robust fit of 
our data over the whole measured temperature range. That $\nu$ follows an activated temperature dependence over a similar 
temperature range to that observed in other materials, as is illustrated in Fig.~\ref{diff}~(c), strongly suggests that up to 
around $250$~K we can assign the change in $\nu$ to lithium diffusion. Above this temperature it is likely that the muons become 
mobile and this results in either a drop in $\Delta$ or $\nu$ as the form of the data changes.

Arrhenius fits to $\nu$ over the thermally activated region allow us to estimate the energy barriers $E_a$ for lithium diffusion, 
which for $x=0.8$, $0.9$, and $1.0$ are $130(10)$, $80(10)$, and $60(10)$~meV respectively. Extrapolating the fits to 300~K for 
comparison with other measured values gives us estimates of the lithium hopping rate at room temperature of $2 \times 
10^6$~s$^{-1}$ ($x=0.8$), $0.8 \times 10^6$~s$^{-1}$ ($x=0.9$), and $0.5 \times 10^6$~s$^{-1}$ ($x=1.0$). (The extrapolation to 
room temperature introduces an error of $\sim 50$~\% in these values whereas the individual points within the measured range have 
errors around $10$~\%.) 

Taking the primary hopping pathway to be along the $b$-axis~\cite{nishimura08} we can further estimate the diffusion constant for 
LiFePO$_4$. The distance travelled in each hop will be $b/2$ and this leads to a diffusion constant $D_{\rm Li} = b^{2}\nu / 4$. 
Given these assumptions we estimate $D_{\rm Li} = 1.9 \times 10^{-9}$~cm$^2$s$^{-1}$ for the $x=0.8$ sample, and $7.6 \times 
10^{-10}$~cm$^2$s$^{-1}$ and $4.6 \times 10^{-10}$~cm$^2$s$^{-1}$ for the $x=0.9$ and $x=1.0$ samples respectively.

We can compare our estimates for the activation barrier and diffusion constant to those derived from other techniques, which are 
summarized in Table~\ref{table}. Most theoretical work and experiments find $E_a$ for lithium diffusion within the range $600$ -- 
$750$~meV,~\cite{delacourt05,ellis06,molenda06,amin08,hoangxx} although there is both theoretical and experimental evidence for 
$E_a \sim 100$ -- $300$~meV.~\cite{takahashi02,shi03,morgan04,wang07} Smaller energy barriers have been suggested for the 
electronic conduction via the hopping of small polarons and it has been argued that the polaron hopping may be correlated with the 
hopping of lithium ions.~\cite{ellis06,maxisch06,hoangxx} The activation energy of $\sim 100$~meV that we observe suggests that 
the hopping process observed by the muons is unlikely to be associated with a barrier as large as $600$~meV and this suggests that 
another, perhaps correlated, process facilitates lithium diffusion at lower temperatures. 

The disparity between measurements of $D_{\rm Li}$ from different techniques is far greater than that seen for $E_a$, with values 
ranging from $10^{-16}$ to $10^{-7}$~cm$^2$s$^{-1}$.~\cite{prosini02,franger02,ellis06,yu07,churikov10} Theoretical 
work~\cite{morgan04} and local measurements, such as M\"{o}ssbauer spectroscopy,~\cite{ellis06,dodd07} seem to give larger values 
of $D_{\rm Li} \sim 10^{-13}$ -- $10^{-7}$~cm$^2$s$^{-1}$ than bulk measurements, which give $D_{\rm Li} \sim 10^{-16}$ -- 
$10^{-10}$~cm$^2$s$^{-1}$ (Ref.~\onlinecite{churikov10}). Our estimate lies within the overlap of these groups. This suggests that 
there is a difference between microscopic and bulk determinations of $D_{\rm Li}$ which could result from the effect of the 
LiFePO$_4$/FePO$_4$ phase boundary motion or mesoscopic barriers to lithium diffusion such as the blocking of diffusion channels 
by Fe$_{\rm Li}$ defects and grain boundaries, the latter accentuated by the habit of crystallites to be platelets with the 
$b$-axis as their shortest dimension.

\section{\label{sec:conc} Conclusion}
We have used $\mu$SR to provide a new window on both the magnetic and diffusive properties of this series of olivine phosphates. 
This has shown how the commensurate-incommensurate phase transition in LiNiPO$_4$ occurs without a discontinuity in the internal 
field at the muon stopping site; how the nature of the fluctuations approaching $T_{\rm N}$ in LiCoPO$_4$ are more 
three-dimensional in the muon time window than those found to be quasi-two dimensional in neutron scattering measurements; and 
that the ordering of LiFePO$_4$ is more conventional than the other two materials studied, though again the three-dimensional 
fluctuations are more evident in determining the behavior of the order parameter approaching $T_{\rm N}$. Our measurements also 
provide a new means of investigating the process of lithium diffusion in Li$_x$FePO$_4$, finding a diffusion constant $D_{\rm Li} 
\sim 10^{-10} - 10^{-9}$~cm$^2$s$^{-1}$ and an energy barrier of $E_a \sim 100$~meV.

Shortly before we submitted this paper Ref.~\onlinecite{sugiyama11} was published reporting analogous measurements of LiFePO$_4$. 
Two oscillating components and a fast relaxing component were observed to extend up to $T_{\rm N}$ suggesting that the higher 
frequency component may originate in a metastable muon site or nearly degenerate muon sites between which the muon hops. The 
high-temperature data were parameterized slightly differently but led to nearly identical values of both $D_{\rm Li}$ and $E_a$.  

\acknowledgments
Parts of this work were performed at the Swiss Muon Source, Paul Scherrer Institute, Villigen, CH and the ISIS Facility, UK. We 
thank A. Amato for experimental assistance and the EPSRC and STFC (UK) for financial support.

\end{document}